\documentclass[twocolumn]{article}
\usepackage[utf8]{inputenc} 
\usepackage[T1]{fontenc}    
\usepackage{hyperref}       
\usepackage{url}            
\usepackage{booktabs}       
\usepackage{amsfonts}       
\usepackage{nicefrac}       
\usepackage{microtype}      
\usepackage{lipsum}
\usepackage{amsmath,comment}
\usepackage{mathtools, cuted}
\usepackage{lipsum}
\usepackage{authblk}

\DeclareMathOperator{\sinc}{sinc}

\title{Far-field intensity signature of sub-wavelength microscopic objects}

\author{Maria Bancerek, Krzysztof M. Czajkowski and Rafa{\l} Koty{\'n}ski\thanks{rafalk@fuw.edu.pl}}
\affil{University of Warsaw, Faculty of Physics\\ Pasteura 5, 02-093 Warsaw, Poland}
  
\begin{document}
\maketitle
\begin{abstract}
Information about microscopic objects with features smaller than the diffraction limit is almost entirely lost in a far-field diffraction image but could be partly recovered with data completition techniques. Any such approach  critically depends on the level of noise. This new path to superresolution has been recently investigated with use of compressed sensing and machine learning.  We demonstrate a two-stage technique based on deconvolution and genetic optimization which enables the recovery of objects with features of 1/10 of the wavelength. We indicate   that l1-norm based optimization in the Fourier domain unrelated to sparsity is more robust to noise than its l2-based counterpart. We also introduce an extremely fast general purpose restricted domain calculation method for Fourier transform based  iterative algorithms operating on sparse data.
\end{abstract}


\maketitle

\section{Introduction}
Superresolving optical microscopy becomes increasingly important in medical and biological applications, in nanotechnology, material science etc. Recent advances in computational imaging~\cite{Mait-aop-10-2-409,Stefanoiu:20} and deep learning~\cite{Barbastathis-Optica-6-921-2019} reopen the question of how much resolution can be enhanced by data completition methods~\cite{Rivenson:Optica-4-1437,Szameit:NatMatt:2012-11-455,Zhu-NatMetods-9-721-2012,Ghosh-arxiv-2005.03595,Roberts:Optica-2016-3-803}.
While scanning near-field optical microscopy~\cite{Near_Field_Optics_book} as well as techniques based on fluorescence microscopy~\cite{Hell:Science-2006-316-1153} allow to reach a deeply sub-wavelength resolution down to the order of several nanometers, the resolution of classical optical imaging is restricted by the Abbe diffraction limit. Fluorescence microscopy gave rise to methods that overcome this limit such as stochastic optical reconstruction microscopy~\cite{Rust:NatMet-2006-3-793}, photo-activated localization microscopy~\cite{Betzig:Science-2006-313-1642} or stimulated emission depletion~\cite{Hell:OE-2008-16-4154,Hell:OL-1994-19-780}. 
These techniques bypass the diffraction limit which  refers to two-point resolution or the width of the point spread function but not to the localization precision of point sources~\cite{Cremer:EurPhysJ-2013-38-281}, whereas optical detection of isolated sub-wavelength objects remains a  part of these methods. By purely optical means the resolution may only insignificantly exceed the Abbe limit~\cite{Sheppard:Micron-2007-38-165,Bechhoefer:ajp-2015-83-22}. Respective methods range from using a high refractive index immersion liquid and a high-numerical aperture (NA>1) objective on which the diffraction limit actually depends, through structured illumination~\cite{Gustaffson:Microscopy-2000-198-82} and deconvolution techniques~\cite{DeconvMicroscopy2005,Liu:APL-81-3143,Latychevskaia:APL-103-2013}, up to development of novel optical set-ups such as for 4PI confocal optics~\cite{Hell:OC-1992-93-277,Hell:NatPhoton-29-381}.

Resolution depends on the use of the available degrees of freedom of the imaging system~\cite{Lukosz:JOSA-1966-56-1463,Lukosz:JOSA-1967-57-932} given in terms of its space-bandwidth product and independent polarization channels~\cite{Sheppard:Optik-2003-113-548,Andrews:Nat-2001-409-316,Sheppard:Micron-2007-38-165}. It may be increased by using \textit{a priori} knowledge about the object combined with signal modulation and reconstruction techniques~\cite{Dekker:JOSA-1997-14-547,Szameit:NatMatt:2012-11-455}. A well known encoding technique that enables to extend the measured spatial spectrum of the specimen by copying high frequency information to lower frequencies is based on the use of gratings~\cite{Lukosz:JOSA-1966-56-1463,Lukosz:JOSA-1967-57-932,Sentenac:PRL-2006-97-243901,Sentenac:PRL-2006-97-243901}. More generally, modulation may involve structured illumination~\cite{Gustaffson:Microscopy-2000-198-82}, speckle pattern projection~\cite{Garcia:OE-2005-13-6073,Shergei:18}, or structured illumination varied on sub-wavelength scale~\cite{Sentenac:PRL-2006-97-243901,liu2007ffo,Narimanov:ACSPhot-2016-3-1090,Ma:ACSNano-2018-12-11316}.
A potential novel promising approach to superresolution is based on superoscillations~\cite{Chen:LSA-2019-8-56,Yuan:LSA-2019-8-2,arXiv:1908.00946-superosc}. Imaging with nanospheres has been also shown to overcome the diffraction limit~\cite{Wang:NatCommun-2011-2-18}. Finally, digital processing with deep neural networks may enhance the spatial resolution of regular microscopic images slightly beyond Abbe's limit~\cite{Rivenson:Optica-4-1437}.

In this paper we make use of far-field intensity pattern obtained under coherent illumination to deduce information on the shape and location of sub-wavelength sized objects. The specimen is placed in slits of a sub-wavelength binary metallic grating. Interference of the Fourier spectra of the object(s) and the grating enhances far-field intensity modulation introduced by the object(s). This approach resembles object recovery in deconvolution microscopy~\cite{DeconvMicroscopy2005}. It is also similar to the interscale mixing microscopy (IMM)~\cite{Inampudi:OE-2015-23-2753,Roberts:Optica-2016-3-803}, although we do not share the opinion~\cite{Roberts:Optica-2016-3-803} that a binary grating nonoverlapping with the object could introduce mixing of the evanescent spectrum of the object into the far-field pattern. In fact this concept is not upheld in~\cite{Ghosh-arxiv-2005.03595}. As compared to IMM, we use a different singal reconstruction method and consider a 2D situation. The far-field diffraction intensity pattern is captured and processed to obtain information on the object. 
We apply a two-step image reconstruction method  for sub-wavelength objects and demonstrate that signal recovery from the far-field diffraction pattern remains feasible in 2D despite the significant drop in the fraction of diffracted light within the diffractive pattern observed when a 2D situation is compared against 1D. Numerical recovery of sub-wavelength objects from their far-field interference signatures is computationally challenging and depends on using further assumptions, for instance on the sparsity of the objects. The overall resolution and amount of details that may be recovered from the non-evanescent field is strongly limited by the signal-to-noise ratio. 

\begin{figure}[htbp]
    \centering
    \includegraphics[width=0.8\linewidth]{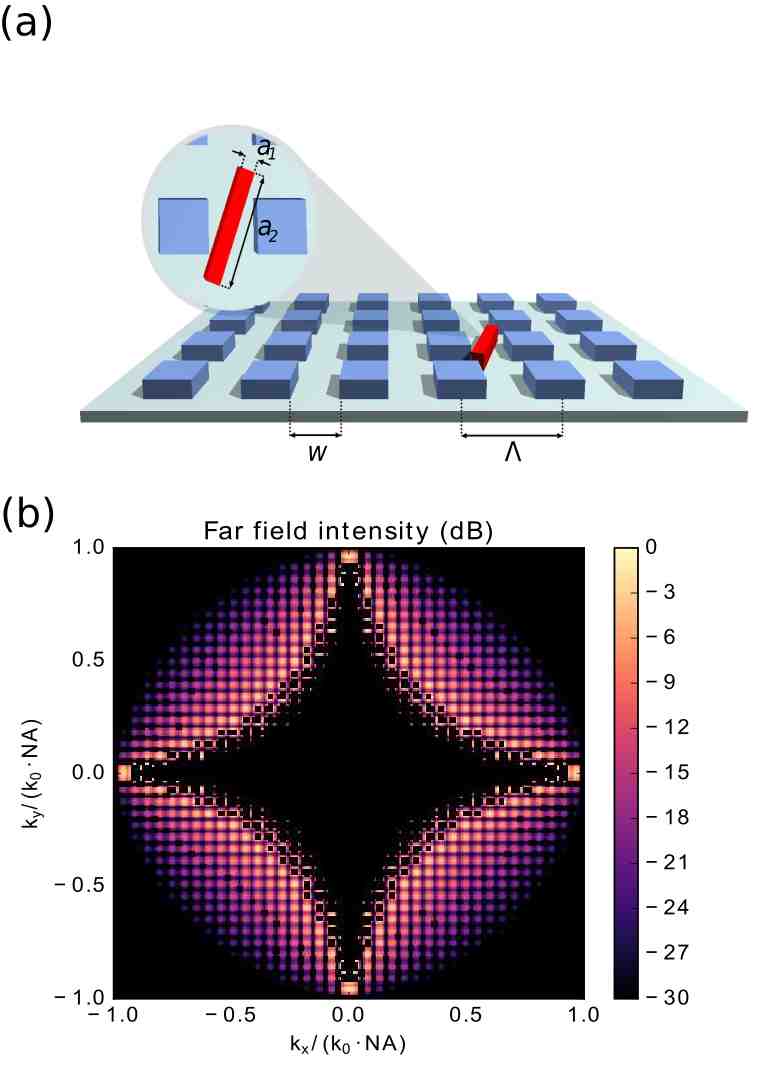}
    \caption{(a)~Subwavelength-sized object (in red) placed in a slit of a 2D binary grating. (b)~Far field intensity expressed in dB (and obtained for an object shown in Fig.~\ref{fig:devonv_ex}(a). The dynamic range of the measurement is limited to 30~dB which results in masking overexposed low-frequency information. }
   \label{fig:schematic}
\end{figure}

\begin{figure}[htbp]
    \centering
    (a)\includegraphics[width=6cm]{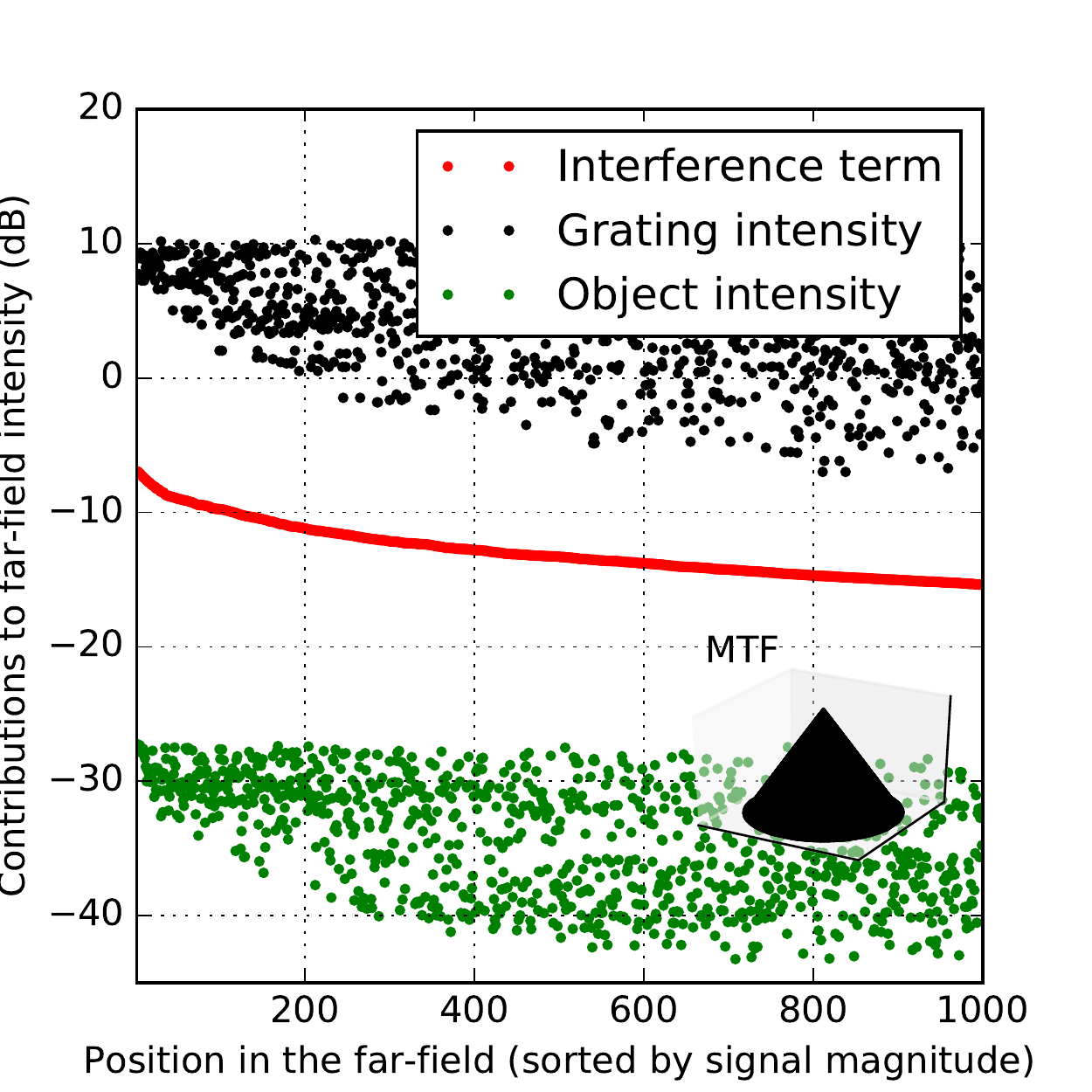}
    (b)\includegraphics[width=6cm]{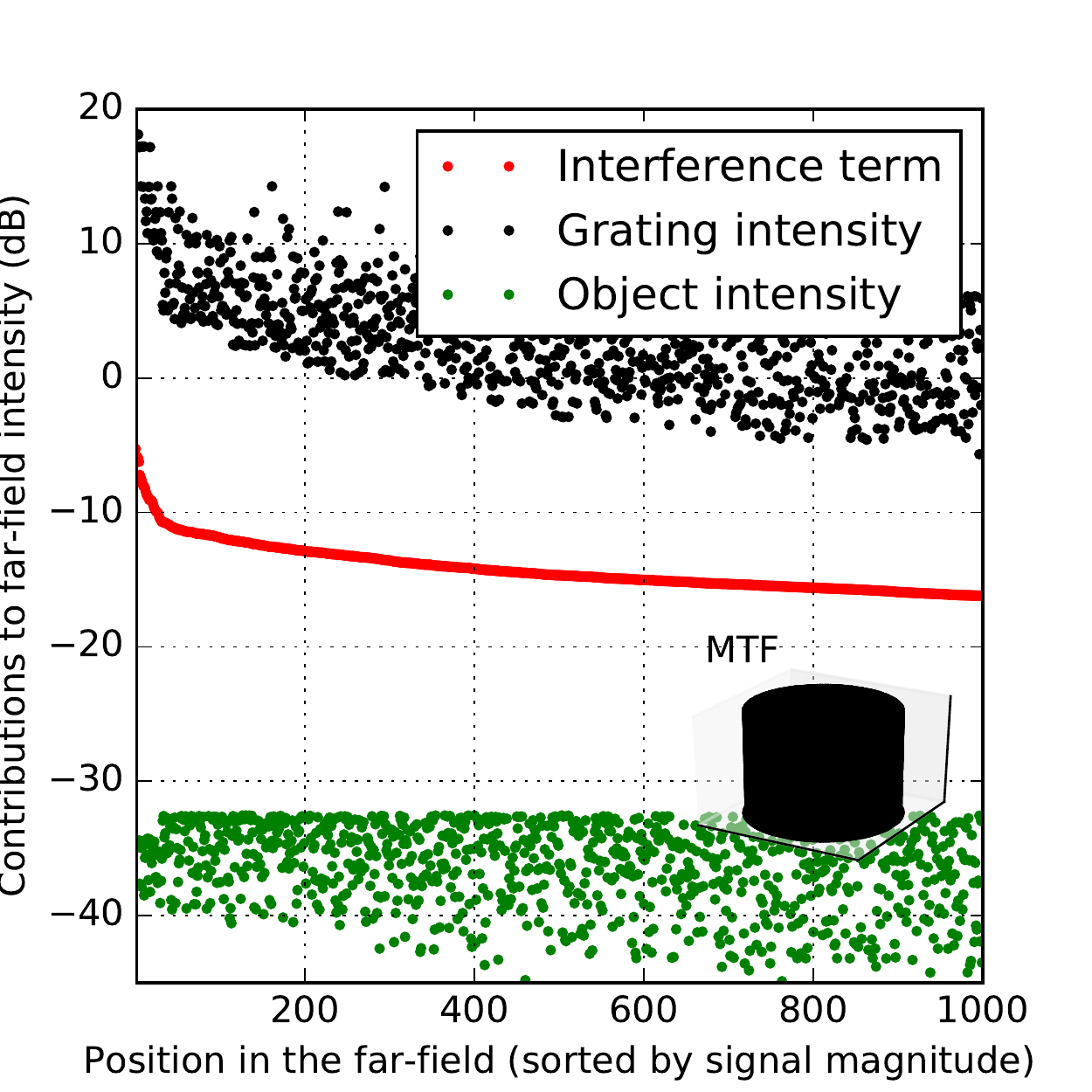}
    \caption{Contributions to the far-field interference pattern at different points of the far-field image (sorted by intensity and normalized by the measured  mean far-field intensity $<I>$). The interference term $I_{int}/<I>$~(in red) is by $2$ to $3$  orders of magnitude larger than the intensity of the object $|\hat O|^2$ (green) and can be measured. (a)~far field determined with a conical modulation transfer function (MTF) of a lens, (b)~far field determined with a binary band-limited circularly-shaped MTF (for MTF-compensated systems or for Fresnel diffraction).}
   \label{fig:enhancement}
\end{figure}

\section{Recovery of sub-wavelength objects from a far-field interference pattern}
Reconstructing an object from the intensity measurement of its  far-field interference pattern is an ambiguous inverse problem. The proposed method consists of two parts. The first stage is derived from the  framework of deconvolution microscopy. Then we apply an original optimization procedure with a genetic algorithm using a criterion evaluated with a restricted-domain Fourier transform.

Figure~\ref{fig:schematic} shows a sample object placed together with a binary grating and introduces respective denotations for the geometric features. The field from the grating interferes with that from the object(s)  placed within the slits of the grating. Intensity distribution is measured in the Fourier plane of the objective. The overall far field intensity is $I(k_x,k_y)\propto |(\hat R(k_x,k_y) + \hat O(k_x,k_y))\cdot \hat H(k_x,k_y)|^2$, where $\hat H(k_x,k_y)$ is the optical transfer function (OTF) and its squared modulus is the modulation transfer function (MTF) of the optical system, $O(x,y)$ is the object field, and $R(x,y)$ is the reference field created by the grating. 
The caret denotes the 2D Fourier transform. OTF is a low-pass function with a cut-off at  $|\textbf{k}|<NA\cdot k_0$, where NA is the numerical aperture and $k_0$ is the wavenumber. When the object $O$ fits in the grating slits, it is not modulated  by the grating $R$ and the far field contains a superposition of the two spectra filtered independently by the same OTF. In effect, the  far-field holds no information about the spatial spectrum of the object above the cut-off and the role of the grating in the measurement is other than to shift the evanescent spectrum below the cut-off. For small objects, the interference term $I_{int}=2Re(\hat R^{*}(k_x,k_y) \cdot \hat O(k_x,k_y))\cdot|H(k_x,k_y)|^2$ present in $I(k_x,k_y)$  carries a lot more energy than the object term $|\hat O(k_x,k_y)|^2$. For instance for 2D square objects and a 2D grating with square masks the reinforcement of the intensity signal due to interference is on the order of the squared ratio of their surfaces and may be substantial. Figure~\ref{fig:enhancement} illustrates this enhancement for a 2D $25\times25$ rectangular grating with $\Lambda=275$~nm, $w/\Lambda=50\%$ at the wavelength of $\lambda=532$~nm and for $NA=1.49$ when the size of the object is $15$~nm. The interference contribution from this deeply sub-wavelength object is clearly a measurable correction to the far-field intensity of the grating. The interference pattern also encodes the phase of $\hat O$ as interference fringes. This justifies the use of the grating in our set-up. At the same time, the interference mechanism enhances noise in a similar way as  object information.

Recording the far field interference pattern $I(k_x,k_y)$ resembles recording a Fourier hologram with a reference beam $\hat R(k_x,k_y)$.  The denotation $\hat R$ underlines the role of a reference beam in the holographic recording played by the far-field diffraction pattern of the grating. Similar as in deconvolution microscopy we want to recover $O(x,y)$ from the interference pattern. For a 2D diffraction  grating $\hat R$ can be written as
\begin{equation}
  \hat R(k_x,k_y)=\hat P(k_x,k_y)\cdot \hat S(k_x,k_y)/ \hat Q(k_x,k_y),\label{eq.grt}
\end{equation}
 where $\hat P(k_x,k_y)$ is the Fourier transform of the elementary cell of the grating, $\hat P=const-w^2 \sinc\left({k_x w}/{2}\right)\sinc\left( {k_y w}/{2}\right)$, and $\hat S/\hat Q$ depends on the distribution of cells in the finite sized grating,
 $\hat S(k_x,k_y)= {\sin\left({ k_x N\Lambda}/{2}\right)}  {\sin\left({ k_y N\Lambda}/{2}\right)}$,
 and $\hat Q(k_x,k_y)= {\sin\left({k_y \Lambda}/{2}\right)\sin\left({k_x \Lambda}/{2}\right)}$. Here $N$, $w$, and $\Lambda$ denote the number of grating periods in each dimension, width of a square grating mask, and the grating pitch, respectively, as indicated in Fig.~\ref{fig:schematic}.
 The object may be approximately recovered with the deconvolution formula
 \begin{equation}
      O^{deconv}(x,y)=\mathcal{F}^{-1} \left\{\frac{I\cdot \hat Q}{\hat P \cdot \hat S \cdot|\hat H|^2}- \frac{\hat P \cdot \hat S}{\hat Q}\right\},\label{eq.rk}
\end{equation}
where the inverse Fourier transform is calculated over part of the domain for which $|\hat P \cdot \hat S \cdot \hat Q \cdot \hat H|>0$. The result is diffraction limited. For a sub-wavelength sized object all fine details are lost, and closely positioned objects can not be resolved, but the locations of the objects may be approximately determined restricting the vast domain that will be examined in further computationally intensive optimization. 
The deconvolution formula~(\ref{eq.rk}) is used by us only to estimate the region of pixels where the object(s) may be located. 

The second stage of the algorithm consists of minimizing an error function dependent on the hypothetical object location and shape using the  actual far-field intensity measurement. We consider both the $\ell_1$ and $\ell_2$ norms to construct the criterion,
\begin{equation}
    F_p(O^{hyp})= \left( \sum_{i\in\Omega} \left|I(k_{x_i},k_{y_i}) -I^{hyp}(\hat O^{hyp}(k_{x_i},k_{y_i})) |^2\right|^p\right)^{1/p},\label{eq.crit}
\end{equation}
where $O^{hyp}$ is the tested hypothesis of the objects shape, and $I^{hyp}=|\hat H(k_{x_i},k_{y_i})|^2\cdot |\hat R(k_{x_i},k_{y_i}) + \hat O^{hyp}(k_{x_i},k_{y_i})|^2$ is the corresponding far-field intensity which may be compared to the actual measurement $I$.

In Eq.~(\ref{eq.crit}), $p=1,2$ decides upon the use of $\ell_1$ or $\ell_2$ norm, and $\Omega$ denotes a subset of spatial frequencies selected adaptively based on signal to noise ratio. Optimization is computationally intensive, and may only be  successful with additional constraints imposed on the object $O(x,y)$. We assume that $O$ is sparse, and is located in the region first estimated in the deconvolution stage. The size of this region considered here is on the order of $1000$ pixels, while the number of spatial frequencies in $\Omega$ is on the order of $10^2- 10^4$ (we used the value of $300$ most of the time). Optimization is further simplified for binary objects of a known and fixed or parametrized shape.

\begin{figure}[htbp]
    \centering
 \includegraphics[width=5.5cm]{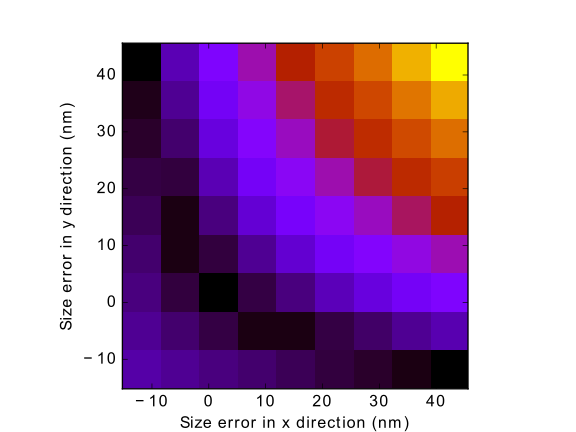}
    \caption{Criterion $F_1$ calculated in the case when two square objects with sizes $53\times 53$~nm and $23\times 23$~nm are located in close vicinity. The tested hypothesis assumes that we know the exact object locations and the size of the larger object. $F_1$ is analyzed as a function of the size of the smaller object. This example shows that although the global minimum appears for the correct hypothesis about the size of the object, other local minima also exist.}
    \label{fig:shape_of_minimum}
\end{figure}

In practice Eq.~(\ref{eq.crit}) has a large number of local minima and can not be minimized with gradient descend methods. In Fig.~\ref{fig:shape_of_minimum} we show a typical shape of $F_p$ near the global minimum, as a function of two parameters corresponding to the possible size of the object when its location is already known. The minimum is broad, and some local minima are also present. The situation becomes a lot more complicated when more degrees of freedom need to be included in optimization. In practice the number of local minima makes optimization difficult. 

For this reason, we have used a genetic algorithm for finding the minimum, which had the additional advantage of the ease to encode additional structural information of the objects. The genetic algorithm was also more robust than Gerchberg-Saxton type iterative optimization.

Finally, it is crucial to minimize the evaluation time of the criterion. Equation~(\ref{eq.crit}) is formulated in the Fourier domain, but the constraints on $O(x,y)$ about object location, sparsity, shape or parametrization can be only easily specified in the image domain. Thus a 2D Fourier transform has to be calculated every time, when we want to calculate the criterion. It would extremely inefficient to work with dense zero-padded matrices with high resolution sampling and use the FFT algorithm for this purpose. Instead we are calculating the discrete Fourier transform directly over small subsets of signal and spectral domains. This approach was faster by two to three orders of magnitude than using FFT from a highly optimized FFTW package included in Matlab. The details of the genetic algorithm are described in the next section and the details of the restricted domain Fourier transform in the Appendix.

\begin{figure}[htbp]
    \centering
    \includegraphics[width=6cm]{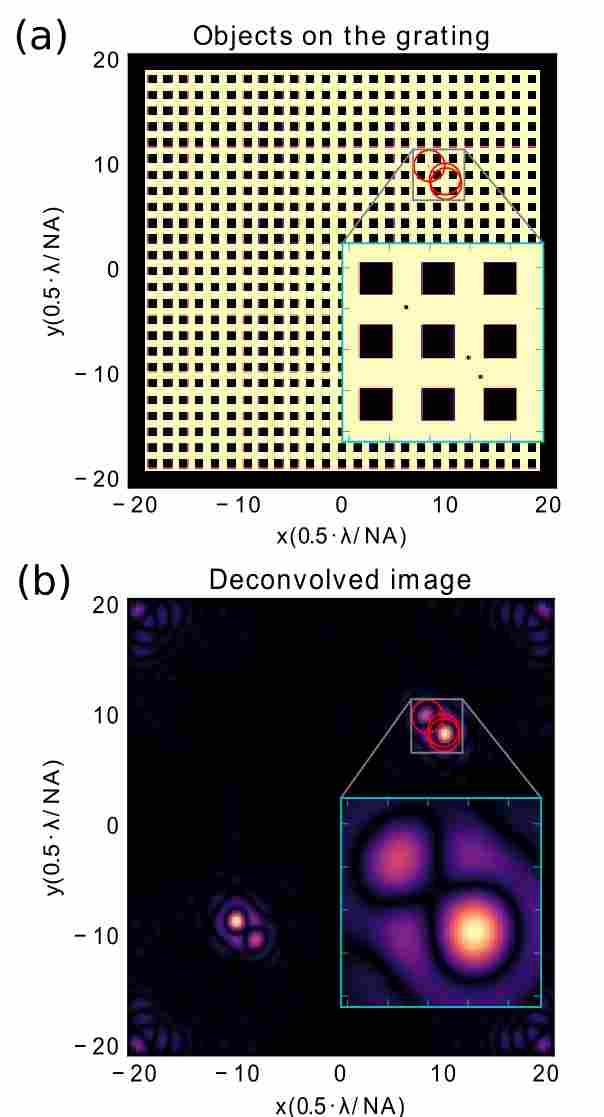}
    \caption{Recovery of sub-wavelength sized objects from a far-field interference pattern by deconvolution: (a)~Three $15$~nm objects placed on the grating (length expressed in the units of diffraction limit). The corresponding far-field interference pattern is shown in shown in Fig.~{\ref{fig:schematic}(b)} (b)~Objects recovered from the far-field interference pattern using Eq.~(\ref{eq.rk}). Isolated objects can be identified and localized but the image is diffraction limited and contains a spurious mirror image. The deconvolved image is the result of the first stage object recovery, which is later refined by numerical optimization beyond the diffraction limit.} 
   \label{fig:devonv_ex}
\end{figure}

 \section{Computational imaging beyond the diffraction limit}
 In order to minimize the cost function defined in Eq.~(\ref{eq.crit}) with respect to positions and sizes of objects we utilize a dedicated genetic algorithm. We consider a population of solutions, each of which contains information about positions and sizes for a set of objects. The objects have a discrete pixelated structure and are located within the area of $\simeq 1000$ pixels selected with the deconvolution formula~(\ref{eq.rk}). The initial population is generated randomly with a uniform distribution of objects over the allowed area.
  Then, in each algorithm iteration, a new population is generated with the better half of the solutions preserved, and the other half regenerated with three genetic operators, which are mutation applied to object position, mutation applied to object size, and crossover. 

The genetic algorithm can be easily adapted to work with parametrized objects. In this work we also analyze arbitrarily oriented line-shaped objects fully characterized by the coordinates of their ends and their width.
  

\begin{figure}[htbp]
\centering
\includegraphics[width=0.8\linewidth]{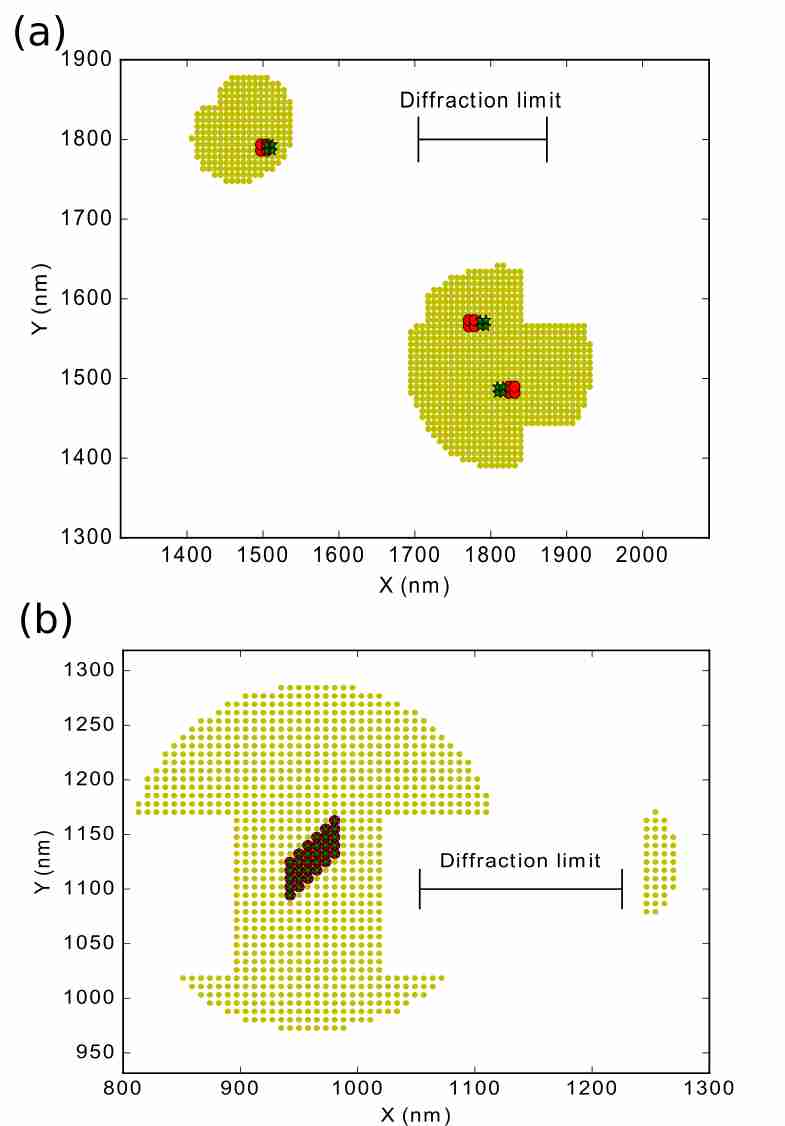}
\caption{Recovery of sub-wavelength sized objects from a far-field interference pattern with a genetic algorithm (red). The potential areas where the objects may be located (marked in yellow) were initially determined by deconvolution The actual object locations are shown in green. (a)~Three $15$~nm objects found approximately ~(See.~Fig.~\ref{fig:devonv_ex}(b.); (b)~single linear object parametrized with $5$ coefficients found exactly. Optimization with the genetic algorithm is also shown in  \textit{Visualisation 1}.}
\label{fig:genetic_ex}
\end{figure}

Figure~\ref{fig:genetic_ex}(a) presents an example of three sub-wavelength objects reconstructed with the genetic algorithm.  The objects and their preliminary localization with deconvolution are shown in Fig.~\ref{fig:devonv_ex}. The algorithm does not always converge to the exact solution but clearly allows to achieve a localization accuracy on the order of $\lambda/10$ and to resolve objects located at distances smaller than the diffraction limit. In Fig.~\ref{fig:genetic_ex}(b) we show a similar example but there is a single linear object parametrized with $5$ coefficients representing the coordinates of the two ends and the thickness. The genetic algorithm is aware of the possibility to parametrize the object and the coefficients are recovered without error.  Optimization with the genetic algorithm is also shown in the supplementary materials (See Visualisation 1).

\begin{figure}[htbp]
    \centering
\includegraphics[width=0.8\linewidth]{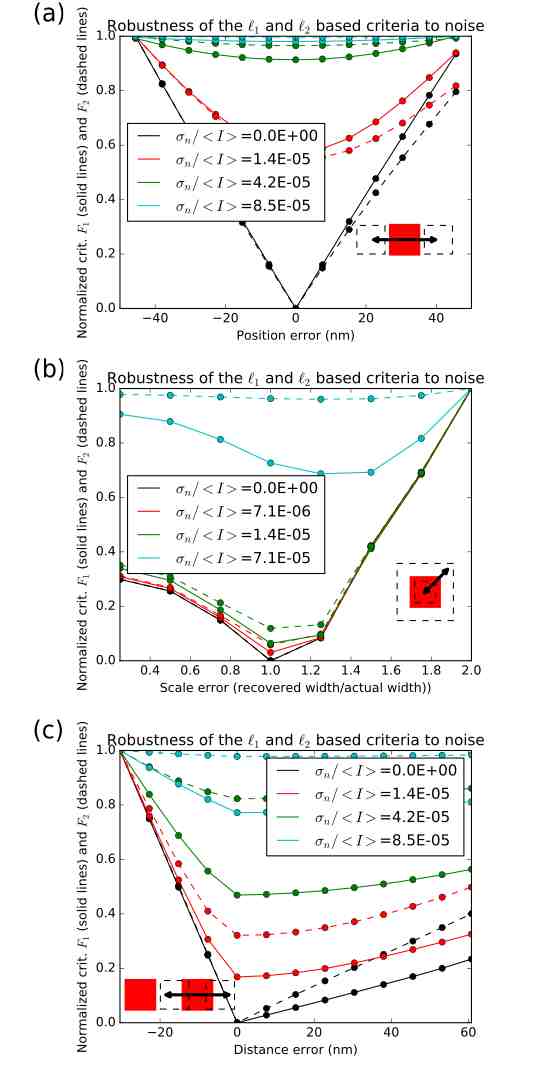}
\caption{Sensitivity of $\ell_1$ and $\ell_2$ norm based criteria to noise and distortions. Criteria $F_1$ and $F_2$ are calculated in the presence of noise when the actual object consists of a single $30$~nm square spot (a),(b) or of two such objects separated by $30$~nm (c). The change of the shape of the minimum with noise indicates that $\ell_1$-based criterion is more robust to noise. (a)~the tested hypothesis assumes a correct size but includes a position error; (b)~the tested hypothesis assumes a correct position and proportions but includes a scale error. c)~the tested hypothesis assumes correct positions and sizes but unknown separation.   }
   \label{fig:noise}
\end{figure}

\begin{figure}[htbp]
    \centering
    \includegraphics[width=\linewidth]{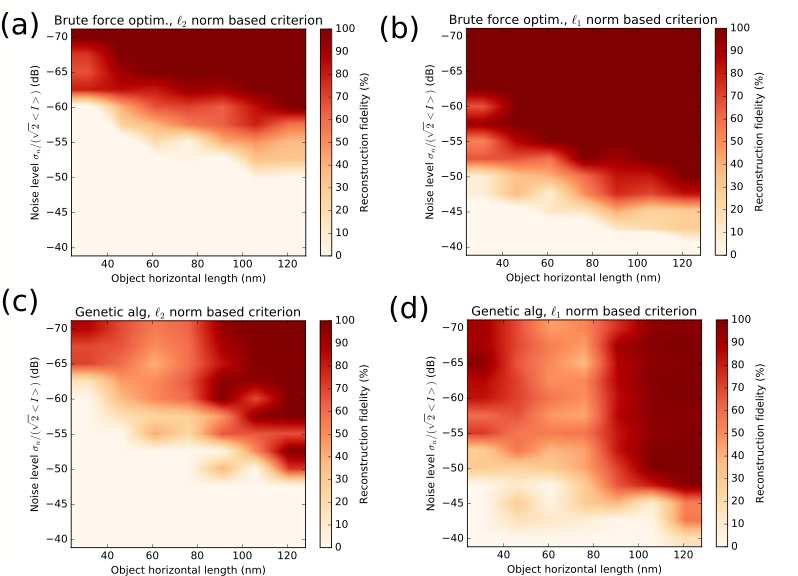}
    \caption{Reconstruction fidelity of a sub-wavelength linear object found from far-field interference pattern in the presence of noise by optimization of the criterion defined in Eq.~(\ref{eq.crit}) with either $p=2$~(a),(c) or with $p=1$~(b),(d). An example of the object is shown in Fig.~\ref{fig:genetic_ex}(b). (a),(b)~results corresponding to a global minimum obtained by brute-force optimization, (c),(d)~results obtained with a genetic algorithm. The object consists of a single $33$~nm-thick line with length varied between $30$ and $120$~nm. Results are extremely sensitive to noise and indicate a better noise robustness obtained with $\ell_1$ norm than with a $\ell_2$ norm.}
   \label{fig:noise1}
\end{figure}

\section{Influence of noise on image reconstruction}
Presence of noise in the far-field intensity measurement has a profound influence on the possibility of object reconstruction and on its quality.  As a measure of noise intensity we will use the ratio of its standard deviation $\sigma_n$ to the intensity of the far-field averaged over the measured region $<I>$ (the same normalization of far-field contributions was used in Fig.~\ref{fig:enhancement}). Noise is complex and affects the complex field additively. Due to the presence of noise, optimization of criteria $F_1$ and $F_2$ is not equivalent, and there is no guarantee that either of them has still a global minimum for the correct hypothesis about the microscopic object shapes and locations.
Criterion $F_1$ based on the $\ell_1$ norm is more robust to noise than $F_2$. We note that this observation can not be simply attributed to linking object sparsity to the value of the~$\ell_1$ norm because the criteria are formulated in the Fourier rather than object domain, and only sparse objects are considered in the genetic algorithm. 
In Fig.~\ref{fig:noise} we compare the two criteria calculated in the presence of noise for the hypothesis about object shape or location near the correct values. The global minimum to the optimization problem becomes shallower with the increasing level of noise. By comparing $\partial F_i/\partial \sigma_n$ with $\partial F_i/\partial \epsilon$, where $\epsilon$ represents location error or size error, we notice that $F_2$ criterion loses sensitivity to these errors much faster than $F_1$ when noise is present. A fundamental question is up to what level of noise $\sigma_n$, the criteria $F_1$ and $F_2$ retain the global minima at the correct location or in its close vicinity. When this level is exceeded, it will no longer be possible to recover the objects, independently of the computational resources available. In practice though, for more complicated problems it may be extremely difficult to find the global minimium, and especially a flat and shallow minimum may be easily omitted by numerical optimization.

We will now focus on localizing a single binary-valued linearly-shaped sub-wavelength sized object positioned on the discrete rectangular pixel grid (such as shown in Fig.~\ref{fig:genetic_ex}(b)). A regular object of this kind could be interesting for data-storage or security applications, where a simple far-field interference measurement  provides more information than is possible to see with an optical microscope at the same wavelength. An object that is fully parametrized by $5$ integer numbers is a convenient test example since for these number of degrees of freedom we are able to compare the operation of an iterative algorithm with a brute-force optimization of criteria $F_1$ and $F_2$ over the entire parameter space. Such a comparison, calculated at various noise levels, is presented in Fig.~\ref{fig:noise1}.
The fidelity of the object reconstruction with values between $0$ and $100\%$ is defined as  
\begin{equation}
    fidelity=max \left(1-\frac{S_{err}}{S_{obj}} ,0\right),
\end{equation}
where $S_{obj}$ represents the actual surface of the objects and $S_{err}$ represents the incorrectly identified object surface (including both omitted and erroneously attributed areas). Fidelity of $100\%$ corresponds to a perfectly identified object, while the fidelity of $0\%$ usually signifies that less than half of the object pixels have been correctly identified.
Two important conclusions may be drawn from the results in Fig.~\ref{fig:noise1}. The first is that the $\ell_1$-norm based optimization is more robust to noise than the $\ell_2$ based optimization. The second is that sensitivity to noise puts a severe limitation to the possibility of sub-wavelength sized object recovery from far-field intensity information. Since we have tested the full possible parameter space for a rather idealistic object parametrization, we do not expect any optimization method, including methods of compressed sensing or deep learning, to overcome these limitations. At the same time, the recovery is feasible, if only the level of signal-to-noise ratio is sufficiently large, which may be potentially achieved by temporal or spatial signal averaging, limiting the aperture, improving experimental stability, or other noise-reduction techniques. 

\section{Conclusion}
We have examined the possibility to recover geometrical information on sub-wavelength sized objects, not limited to their location, from a far-field interference pattern obtained under coherent illumination. The far-field signature of microscopic objects considered by us does not include spatial frequencies beyond the cut-off, i.e. corresponding to evanescent waves or spatial frequencies lost in classical imaging optics with an objective having a given numerical aperture (with $NA=1.49$  assumed in the presented results). We have proposed a two-step object recovery algorithm, with the first diffraction-limited step based on deconvolution, and the second based on numerical optimization. This second step involves the use of a genetic algorithm and a restricted-domain Fourier transform (described in the Appendix) the purpose of which is to speed-up calculations of the far-field for sparse objects when only a limited part of the Fourier coefficients are required. Overall, the method makes it possible to recover sub-wavelength sized objects with sizes on the order of $\lambda/10$ from far-field information, although for more complicated scenes, the method becomes computationally intensive and does not always converge to the correct result. Sub-wavelength object recovery is only possible at a very low level of noise. 
Even with a known object parametrization, a small level of noise introduces false minima to the cost functions making the true hypothesis impossible to distinguish from alternatives. This limitation concerns object recovery with both $\ell_1$-norm and $\ell_2$-based criterion functions and is unlikely to be mitigated by the use of sophisticated optimization frameworks such as compressive sensing or deep learning. At the same time, the robustness of $\ell_1$-norm based criterion to noise is considerably better than that of the $\ell_2$-based counterpart. Compressive sensing heavily relies on $\ell_1$ norm as a measure of signal sparsity which leads to convex computationally tractable  optimization criteria. Perhaps noise robustness properties of the $\ell_1$ norm  are in effect underestimated. Our optimization of $\ell_1$-norm applied to dense Fourier domain information is probably original.

It is unlikely that information recovery from a far field signature could become a significant alternative to existing superresolving wide-field microscopic techniques. Prospect applications may be connected with security purposes, where a specific microscopic object impossible to be analyzed by classical optical microscopy could be detected, localized and verified by the examination of the far-field diffraction pattern. Another area of interest is high density data storage where the geometry of sub-wavelength sized features could be used to enhance the amount of information readable without shortening the laser  wavelength.

\textbf{Funding  Information}\\ Narodowe Centrum Nauki (NCN) (2017/27/B/ST7/00885).

\section*{Appendix}
It is hard to overestimate the role of Discrete Fourier Transform (DFT) in numerical techniques used in optics, especially in areas related to Fresnel and Fraunhofer diffraction~\cite{Mas-OC-164-233-1999} that depend on the iterative use of 2D DFT. DFT is  heavily used in the multiple projections methods applied in the design of diffractive elements, in deconvolution and in phase retrieval algorithms~\cite{Fienup:s,Bauschke:02,Momey:19,Latychevskaia:15}, as well as for pulse retrieval in frequency resolved optical gating etc. The best known Fast Fourier Transform (FFT) algorithm~\cite{FFT1965} is certainly most commonly used. However the need for a uniform sampling in both signal and frequency domains at the same time implies operating on large zero-padded arrays, which is inefficient. Also shifted Fresnel transform~\cite{Muffoletto:07}, as well as techniques involving nonuniform sampling with Bluestein's algorithm~\cite{Hu:LSI-9-119-2020} do not allow to operate on arbitrarily selected parts of signal and Fourier domain. 

This Appendix describes the way we calculate the 2D discrete Fourier transform (DFT)  over restricted parts of image and spectral domains. Sample code is available at~\cite{rdft}. We actually apply directly the 2D DFT definition to sparse matrices and express the result as a simple matrix-vector product, where the  matrix is dynamically constructed from a set of values precalculated in advance. This allows us to calculate efficiently the 2D DFT of arbitrarily distributed sparse objects at arbitrarily selected frequencies. We consider this result as an important coding detail rather than a novel method, which however may give a huge calculation speedup to iterative sparse diffraction calculations. Using restricted domain DFT we were able to decrease the calculation time by $2$ to $3$ orders of magnitude, as compared to the use of ordinary FFT with full matrices.

The 2D DFT of a vector $v$ (indexed with two indices) is
\begin{equation}
    \hat v_{k_x,k_y}\propto \sum_{m_x=0}^{N_x-1} \sum_{m_y=0}^{N_y-1}  v_{m_x,m_y}\cdot \exp\left(2\pi i \left(\frac{m_x k_x}{N_x} +\frac{m_y k_y}{N_y} \right)\right).
\end{equation}
Let $N$ be the smallest common multiple of $N_x$ and $N_y$, and $N=P_x\cdot N_x=P_y\cdot N_y$. Let $\phi(l)=\exp(2\pi i l/N)$ with $l=0..N-1$.
Now the 2D DFT may be written as,
\begin{equation}
    \hat v_{k_x,k_y}\propto \sum_{m_x=0}^{N_x-1} \sum_{m_y=0}^{N_y-1}  v_{m_x,m_y}\cdot \phi\left((m_x P_x k_x+m_y P_y k_y)mod_N\right),
\end{equation}
where $mod$ is the modulo operation.
The same can be written with a single index assuming column ordering of the images, $k=k_x N_y+k_y$, $m=m_x N_y+m_y$,
\begin{equation}
    \hat v_{k\in\Omega}\propto \sum_{m\in\Theta}   v_m\cdot \phi\left((m_x(m)\cdot P_x\cdot k_x(k)+m_y(m)\cdot P_y\cdot k_y(k))mod_N\right).\label{eq.restrdom_ft}
\end{equation}
Assume that $v$ is a sparse image with only $M<<N_x\cdot N_y$ elements making a set $\Theta$, and we need to determine only $K<<N_x\cdot N_y$ Fourier coefficients that make a set $\Omega$. Since the $N$ elements of $\phi$ may be precalculated in advance, according to Eq.~(\ref{eq.restrdom_ft}) we only need to perform $K\cdot M$ multiplications in a single matrix by vector multiplication. A full 2D FFT would require $N_y\cdot N_x \cdot lg_2 (N_x N_y)$  in the best case when $N_x$ and $N_y$ are powers of $2$. Additional savings in computational time come from a much lower memory usage and the possibility of storing sparse matrices in CPU cache.

\end{document}